\documentstyle[12pt]{article}
\begin{document}
\renewcommand{\Re}{{\rm Re \,}}
\renewcommand{\Im}{{\rm Im \,}}
\newcommand{\Tr}{{\rm Tr \,}}
\newcommand{\beq}{\begin{equation}}
\newcommand{\eeq}[1]{\label{#1}\end{equation}}
\newcommand{\bea}{\begin{eqnarray}}
\newcommand{\eea}[1]{\label{#1}\end{eqnarray}}
\newcommand{\dirac}{/\!\!\!\partial}
\newcommand{\Dirac}{/\!\!\!\!D}
\begin{titlepage}
\begin{center}
\hfill hep-th/9510074,  NYU-TH-95/10/02
\vskip .01in \hfill CERN-TH/95-268, UCLA/95/TEP/33
\vskip .4in
{\large\bf Minimal Higgs Branch for the Breaking of Half of the
Supersymmetries in N=2 Supergravity}
\end{center}
\vskip .4in
\begin{center}
{\large Sergio Ferrara}$^a$,
{\large Luciano Girardello}$^b$ and
{\large Massimo Porrati}$^c$,
\vskip .1in
$a$ Theory Division, CERN, CH-1211 Geneva 23, Switzerland
\vskip .05in
$b$ Theory Division, CERN, CH-1211 Geneva 23, Switzerland~\footnotemark
\footnotetext{On leave from
Dipartimento di Fisica, Universit\`a di Milano, via Celoria 16,
20133 Milano, Italy.}
\vskip .05in
$c$ Department of Physics, NYU, 4 Washington Pl.,
New York NY 10003, USA~\footnotemark
\footnotetext{On leave from INFN, Sez. di Pisa, 56100 Pisa, Italy.}
\end{center}
\vskip .4in
\begin{center} {\bf ABSTRACT} \end{center}
\begin{quotation}
\noindent
It is shown that the minimal Higgs sector of a generic N=2 supergravity
theory with unbroken N=1 supersymmetry must contain a Higgs hypermultiplet
and a vector multiplet. When the multiplets parametrize the quaternionic
manifold $SO(4,1)/SO(4)$, and the special K\"ahler manifold $SU(1,1)/U(1)$,
respectively, a vanishing vacuum energy with a sliding massive spin 3/2
multiplet is obtained. Potential applications to N=2 low energy effective
actions of superstrings are briefly discussed.
\end{quotation}
\vfill
\end{titlepage}
\noindent
The field theoretical analysis of the Higgs and the super-Higgs mechanisms has
already proven itself in the past to be a powerful tool to analyse phenomena
that may occur in string theory.

Recently, conifold transitions in type II strings compactified on
Calabi-Yau manifolds~\cite{0} have been described by Greene, Morrison and
Strominger~\cite{1}
as a Higgs mechanism in which $p$ hypermultiplets are ``eaten up'' by $p$
$U(1)$ massless vector multiplets, which become $p$ massive,
long vector multiplets~\cite{2}
(i.e. multiplets with vanishing central charge), each with spin content
$(1,1/2(4),0(5))$.
This Higgs branch is a particular case of a phenomenon, previously noted in
the context of supersymmetric gauge theories~\cite{3}, where, generically,
VEV's of hypermultiplets can change the rank of the (unbroken) gauge group.
This is is contrast with the Coulomb phase in which VEV's of vector
multiplets do not change the rank of the gauge group.

Purpose of the present work is to investigate a much richer structure which
emerges in supergravity theories,  in which these branches can induce
supersymmetry breaking together with gauge symmetry breaking.

The new phenomenon which emerges here is that N=2 Fayet-Iliopoulos
terms~\cite{4} can break all or half~\cite{6} of the
supersymmetries depending on whether charged hypermultiplets exist
in the theory~\cite{6}, which couple both to the graviphoton and to the
matter vector multiplets. Partial supersymmetry breaking is also possible
with vanishing vacuum energy~\cite{8}.

Suppose at first that hypermultiplets are not present, but that Fayet
Iliopoulos terms are.
Furthermore, choose an Abelian gauge group $U(1)^{n_V+1}$, where $n_V$ is the
number of matter vector hypermultiplets. Then the Fayet-Iliopoulos term
corresponds to a constant gauge prepotential~\cite{7}
\beq
{\cal P}_\Lambda^x=\xi_\Lambda^x,
\eeq{1}
such that:
\beq
(\vec{\xi}_\Lambda \wedge \vec{\xi}_\Sigma)^z=
\epsilon^{xyz}{\cal P}_\Lambda^x {\cal P}_\Sigma^y=0
\eeq{2}
The vacuum energy, in the notations of reference~\cite{7},
is given by the formula
\beq
V(z,\bar{z},\xi^x_\Lambda)=U^{\Lambda\Sigma}\xi_\Lambda^x\xi_\Sigma^x
-3 \bar{L}\cdot \xi^x L \cdot \xi^x.
\eeq{3}
Here $z$ denotes the scalars of the vector-multiplet manifold,
whose metric is $g_{i\bar{\jmath}}$, while
\beq
U^{\Lambda\Sigma}=(\partial_i
+{1\over 2}\partial_iK)L^\Lambda
g^{i\bar{\jmath}}(\partial_{\bar{\jmath}} +{1\over
2}\partial_{\bar{\jmath}}K)\bar{L}^\Sigma, \;\;\; L^\Lambda=e^{K/2}X^\Lambda.
\eeq{3'}
The special geometry data $X^\Lambda$, $K$ are defined below, in the
parargraph after eq.~(\ref{9}).
In reference~\cite{4} it was shown that for a particular choice of the
prepotential, $F(X^\Lambda)$,
it is possible to have $V\equiv 0$. However, since $\langle z \rangle $ is
$SU(2)$ invariant, it follows that both gravitini have the same (sliding)
mass~\cite{6}. Therefore, this example breaks all supersymmetries, while the
$U(1)^{N_V+1}$ symmetry is unbroken. It thus corresponds to the Coulomb phase.
In this case gauginos get masses proportional to the gravitino mass.

A more interesting situation arises
when the theory is not in the Coulomb phase,
but rather in the Higgs phase. As in rigid supersymmetry, this can only occur
if matter hypermultiplets are present. The new phenomenon that we want to
emphasize here is that the hypermultiplets not only can give masses to the
$U(1)^{n_V+1}$ gauge bosons, but can also break half of the supersymmetries
rather than all of them, in the presence of Fayet-Iliopoulos terms~\cite{8}.
Interestingly enough this is a phenomenon that has no analog in rigid
supersymmetric theories (whether or not they are renormalizable) simply
because, as pointed out by Witten~\cite{9}, in rigid theories the supersymmetry
algebra implies that if one supersymmetry is broken, then the vacuum energy
is strictly positive, implying that {\em all} supersymmetries are indeed
broken. In supergravity this is circumvented because the supergravity Ward
identities read~\cite{10}
\beq
\delta_A\psi_L^i\delta^B\psi_R^j{\cal Z}_{ij}-3{\cal M}_{AC}\bar{\cal M}^{CB}
=V\delta_A^B,
\eeq{4}
Where the $\delta_A\psi^i$ denote the shift, under the $A$-th
supersymmetry, of the spin
one-half fermions, while ${\cal M}_{AB}$ is the gravitino mass matrix and
${\cal Z}_{ij}$ is the kinetic term of the fermions.
These identities show that even when $V=0$, one may still have, say
\beq
\delta_1\psi^i_L\delta^1\psi^j_R{\cal Z}_{ij}=3{\cal M}_{1C}{\cal M}^{C1}=0,
\eeq{5}
but instead
\beq
\delta_2\psi^i_L\delta^2\psi^j_R{\cal Z}_{ij}=3{\cal M}_{2C}{\cal M}^{C2}\neq
0.
\eeq{6}
In N=2, this corresponds to breaking half of the supersymmetries
(N=1 unbroken), at zero cosmological constant.

A model that realizes such a situation cannot be obtained from the
Lagrangian of De Wit, Lauwers and Van Proeyen~\cite{11}, as it was proven
in~\cite{11a}. On the other hand, that
is not the most general N=2 Lagrangian. It uses, in fact, a symplectic basis
in which a prepotential $F(X)$ exists for the vector multiplets.
In reference~\cite{12}, it was shown that this is not generally true, and that
a more general formulation of N=2 supergravity exists, that never makes use of
the prepotential function.

The minimal model that exhibit partial breaking of $N=2$ supersymmetry to
$N=1$ with zero cosmological constant contains a charged hypermultiplet, whose
scalars parametrize the quaternionic manifold $SO(4,1)/SO(4)$, coupled to a
vector multiplet, whose scalars parametrize the K\"ahler manifold
$SU(1,1)/U(1)$~\footnotemark.
\footnotetext{This model was constructed in~\cite{8} by performing a
singular limit on a model constructed within the framework of the tensor
calculus of ref.~\cite{11}.}
The latter is formulated in a symplectic basis in which no prepotential
exists.

Note that the presence of both a hypermultiplet and a vector multiplet is
needed~\cite{5} since, when N=2 is broken to N=1, the N=1 multiplet containing
the massive spin-3/2 field has spin content $(3/2,1,1,1/2)$. Both the
graviphoton and the matter vector become massive, together with one of the
gravitini; in other words, this is a Higgs and super-Higgs phase.
The spectrum of this theory contains, besides the massive spin-3/2 N=1
multiplet, two massless chiral multiplets with sliding fields, since the
vacuum energy vanishes.

The model is determined by the geometry of the hypermultiplet quaternionic
manifold and the geometry of the vector-multiplet manifold,
together with the the ``D-term'' prepotentials ${\cal
P}_\Lambda^x$~\cite{7}.

Let us denote the quaternionic coordinates of the hypermultiplet manifold
by $b^u$, $u=0,1,2,3$.
The quaternionic geometry is determined
by a triplet of quaternionic potentials,
$\Omega^x=\Omega^x_{uv}db^u\wedge db^v$, $x=1,2,3$,
which are the field strenght of an $SU(2)$ connection
$\omega^x=\omega^x_u db^u$: $\Omega^x=d\omega^x
+(1/2)\epsilon^{xyz}\omega^y\wedge \omega^z$.
In our case, the quaternionic manifold is $SO(4,1)/SO(4)$, and these
quantities read:
\beq
\omega^x_u={1\over b^0}\delta^x_u, \;\;\; \Omega^x_{0u}=-{1\over
2{b^0}^2}\delta^x_u, \;\;\; \Omega^x_{yz}={1\over 2
{b^0}^2}\epsilon^{xyz},\;\; x,y,z=1,2,3.
\eeq{7}
The prepotentials $\Omega^x$ determine the quaternionic metric $h_{uv}$
by the identity~\cite{7}
\beq
h^{st}\Omega^x_{us}\Omega^y_{tv}=-\delta^{xy} h_{uv}
-\epsilon^{xyz}\Omega^z_{uv}.
\eeq{8}
In our case this equation gives $h_{uv}=(1/2{b^0}^2)\delta_{uv}$.
To write the fermion shifts one also needs the symplectic vielbein
${\cal U}^{\alpha A}_u db^u$, $\alpha,A=1,2$~\cite{7}.
In our case the vielbein reads:
\beq
{\cal U}^{\alpha A}={1\over 2b^0} \epsilon^{\alpha\beta}(db^0
-i\sigma^xdb^x)_\beta^{\; A},
\eeq{9}
where $\sigma^x$ are the standard Pauli matrices.

The special geometry of the manifold of the vector multiplets is
determined in general by giving $2n_V+2$ holomorphic sections~\cite{12}
$X^\Lambda(z), F_\Lambda(z)$, in terms of which the K\"ahler potential reads
\beq
K=-\log i(\bar{X}^\Lambda F_\Lambda   -X^\Lambda \bar{F}_\Lambda).
\eeq{10}
Here the manifold is $SU(1,1)/U(1)$, $\Lambda=0,1$,
there is a single holomorphic coordinate $z$, and our
choice of holomorphic sections is
\beq
X^0(z)=-{1\over 2},\;\; X^1(z)={i\over 2},\;\; F_0=iz,\;\; F_1=z.
\eeq{11}
This choice gives rise to the K\"ahler potential
\beq
K=-\log (z+ \bar{z}),
\eeq{12}
and thus to the metric $g_{z\bar{z}}=1/(z+\bar{z})^2$.
It is important to remark that our choice of holomorphic sections is such
that no prepotential $F(X^\Lambda)$ exists~\cite{12}~\footnotemark.
\footnotetext{One can find these sections by the symplectic
transformation $X^1\rightarrow -F_1$, $F_1\rightarrow X^1$ of the basis
specified by the prepotential
$F(X^\Lambda)=iX^0X^1$, which reads, explicitly,
$X^\Lambda, $$F_\Lambda=\partial F /\partial X^\Lambda$.}

Any global symmetry of the hypermultiplet manifold can be gauged. If the
corresponding Killing vectors are $k_\Lambda^u$, the gauge covariant
derivative is $D_{\mu}b^u=\partial_{\mu}b^u + A_{\mu}^\Lambda k_\Lambda^u$.
In our case the gauge group is $U(1)^2$, where one of the $U(1)$ factors
comes from the N=2 graviphoton, and the other from the matter vector.
Therefore, we need two commuting Killing
vectors. Since the metric of our quaternionic manifold is
$\delta_{uv}(1/2{b^0}^2)$, the manifold is symmetric under
arbitrary constant translation of the coordinates $b^1,b^2,b^3$.
Thus, we can for instance choose to gauge the translations along $b^1$ with
the graviphoton, and the translations along $b^2$ with the matter vector.
The corresponding Killing vectors are
\beq
k^u_0=g\delta^{u1},\;\;\; k^u_1=g'\delta^{u2},
\eeq{13}
where $g$ and $g'$ are arbitrary constants (the gauge couplings of the two
$U(1)$'s).
The Killing vectors of a quaternionic manifold are derived from a triplet
of ``D-term'' prepotentials, ${\cal P}_\Lambda^x$ by the equation~\cite{7}
\beq
k_\Lambda^u={1\over 6} \sum_{x=1}^3 h^{vw}\nabla_v{\cal
P}^x\Omega^x_{wt}h^{tu}.
\eeq{14}
In our case one has
\beq
{\cal P}^x_0=g{1\over b^0}\delta^{x1},\;\;\; {\cal P}^x_1=g'{1\over
b^0}\delta^{x2}.
\eeq{15}
It is easily checked that the ``quaternionic Poisson braket~\cite{7}''
of these prepotentials is zero, as it must be for an Abelian gauge group
\beq
\{ {\cal P}_0,{\cal P}_1\}^x \equiv \Omega^x_{uv}k^u_0k^v_1 -{1\over
2}\epsilon^{xyz}{\cal P}_0^y {\cal P}_1^z=0 .
\eeq{16}
The relation between our prepotentials and Killing vectors can be
summarized by the following formula
\beq
{\cal P}_\Lambda^x={1\over b^0}k^x_\Lambda.
\eeq{17}

At this point, we have determined all quantities necessary to write the
fermion shifts. The formulae of reference~\cite{7} give the following
expression for the (constant part of the) gaugino shift
\beq
\delta \lambda^{\bar{z}}_A=
-ig^{z\bar{z}}(\sigma^x)_A^{\; C} \epsilon_{BC}{\cal
P}^x_\Lambda e^{K/2}(\partial_z + \partial_z K)X^\Lambda(z)\eta^B
\equiv W^{\bar{z}}_{AB}\eta^B.
\eeq{18}
Here $\eta^A$ is the N=2 supersymmetry parameter.

The shifts of the hyperini is instead
\beq
\delta \zeta^\alpha = -2 \epsilon_{AB}{\cal U}^{\alpha B}_u k^u_\Lambda
e^{K/2}X^\Lambda(z)\eta^A \equiv {\cal N}_A^\alpha \eta^A.
\eeq{19}

Finally the gravitino shift reads
\beq
\delta \psi_{A\, \mu}= {i\over 2} (\sigma^x)_A^{\; C} \epsilon_{BC} {\cal
P}^x_\Lambda e^{K/2}X^\Lambda(z)\gamma_{\mu}\eta^B\equiv
iS_{AB}\gamma_{\mu}\eta^B.
\eeq{20}
By substituting into these formulas the explicit expressions we obtained for
all quantities involved, we find that all fermionic shifts are proportional
to a single matrix:
\beq
W^{\bar{z}}_{AB}=-i(z+\bar{z})^{1/2}{1\over b^0}X_{AB}, \;\;
{\cal N}_A^\alpha = -i(z+\bar{z})^{-1/2}{1\over b^0}
\epsilon^{\alpha\beta}X_{\beta
A},\;\; S_{AB}=-{1\over 2}(z+\bar{z})^{-1/2}{1\over b^0}X_{AB},
\eeq{21}
where
\beq
X_{AB}=-{g\over 2}(\sigma^1)_A^{\;C}\epsilon_{CB}
+i{g'\over 2}(\sigma^2)_A^{\;C}\epsilon_{CB}=\left(\begin{array}{ll} {g' -g
\over 2} & 0 \\ 0 & {g' + g \over 2} \end{array} \right).
\eeq{22}
In these normalizations, the Ward identity relating the scalar potential
to the fermionic shifts reads
\beq
\delta_{AB} V = -12 (S_{AC})^*S_{CB} +
g_{z\bar{z}}(W^{\bar{z}}_{AC})^*W^{\bar{z}}_{CB} + 2 ({\cal
N}_A^\alpha)^*{\cal N}_B^\alpha.
\eeq{23}
Upon substituting eq.~(\ref{22}), we find that this formula gives
$V=0$ identically, for {\em any} value of $g$ and $g'$. The model has always a
flat potential, and sliding VEVs for the scalar fields $z$ and $b^u$. The
gravitino mass matrix is equal to $2S_{AB}$ (compare eq.~(\ref{4}) with
eq.~(\ref{23})); thus, the ratio of the mass of the two gravitini is
independent on the scalar VEV and equal to $|(g+g')/(g-g')|$. When the gauge
coupling of the two $U(1)$'s are equal in magnitude $g=\pm g'$ (and nonzero),
one of the two gravitini is massless, and N=1 supersymmetry is unbroken.
Obviously, all the fermionic shifts along the unbroken supersymmetry
generator vanish.

It is apparent that the model presented here describes the minimal sector
responsible for the breaking of half of the supersymmetries. It is thus
conceivable that its Lagrangian would also provide a model-independent
description of the interactions of the half-supersymmetry breaking sector of a
very large class of interesting theories.

We may wonder whether phenomena such as have been just described here may occur
in string theory. If the N=2 theory under consideration is coming from a
type IIA theory, then the vectors are R-R states, and the hypermultiplets
carry R-R charges~\cite{1,13}. On the other hand, the breaking of half
of the supersymmetries is only possible if a Fayet-Iliopoulos term is
introduced. In our case the prepotentials ${\cal P}^x_\Lambda$ are
Fayet-Iliopoulos terms since, as shown by eq.~(\ref{15}), they are
independent of the vector multiplets and they are always nonzero at any
point on the hypermultiplet manifold.

It is interesting to remark that, as recently noted in ref.~\cite{15}, a
kind of Fayet-Iliopoulos term was introduced by Romans~\cite{14}
in type IIA supergravity. It induces
an anti-Higgs mechanism for a $U(1)$ 10-D vector field, which is eaten by
the $b_{\mu\nu}$ tensor, that thus becomes massive.
Ref.~\cite{15} also discusses a 10-D form in type II theory which induces a
supersymmetry breaking in string theory.
It is plausible that the mechanism
discussed in this paper, or a generalization thereof, may find applications
in the understanding of non-perturbative phenomena in superstring dynamics.
\vskip .2in
\noindent
Acknowledgements
\vskip .1in
\noindent
S.F. is supported in part by DOE under grant DE-FGO3-91ER40662, Task C, and by
EEC Science Program SC1*-CI92-0789. L.G. is supported in part by
Ministero dell' Universit\`a e della Ricerca Scientifica e Tecnologica,
by INFN, and by EEC Science Programs SC1*-CI92-0789 and CHRX-CT92-0035.
M.P. is supported in part by NSF under grant PHY-9318781.

\end{document}